\documentclass[reprint,amsmath,amssymb,aps,pra]{revtex4-1}

\usepackage{graphicx}
\usepackage{dcolumn}
\usepackage{bm}
\usepackage{hyperref}

\usepackage{xcolor}
\newcommand \David[1] {\bgroup\noindent[\textcolor{blue}{\textbf{David}: #1}]\egroup\ignorespacesafterend}
\newcommand \MZ[1] {\bgroup\noindent[\textcolor{red}{\textbf{MZ}: #1}]\egroup\ignorespacesafterend}
\newcommand{\figref}[1]{Fig.~\ref{fig:#1}}

\newcommand{\Br}{{\boldsymbol{\mathnormal r}}}

\begin{document}

\preprint{APS/123-QED}

\title{Avalanche behavior in creep failure of disordered materials}

\author{D. F. Castellanos}
\affiliation{
	Institute of Materials Simulation, University of Erlangen-N\"urnberg, Dr.-Mack-Stra{\ss}e 77, 90762 F\"urth, Germany
}

\author{M. Zaiser}
\affiliation{
	Institute of Materials Simulation, University of Erlangen-N\"urnberg, Dr.-Mack-Stra{\ss}e 77, 90762 F\"urth, Germany
}
\affiliation{
	School of Mechanics and Engineering, Southwest Jiaotong University, Chengdu 610031, China  
}

\date{\today}

\begin{abstract}
We present a mesoscale elastoplastic model of creep in disordered materials which considers temperature-dependent stochastic activation of localized deformation events which are mutually coupled by internal stresses, leading to collective avalanche dynamics. We generalize this stochastic plasticity model by introducing damage in terms of a local strength that decreases, on statistical average, with increasing local plastic strain. As a consequence the model captures failure in terms of strain localization in a catastrophic shear band concomitant with a finite-time singularity of the creep rate. The statistics of avalanches in the run-up to failure is characterized by a decreasing avalanche exponent $\tau$ that, at failure, approaches the value $\tau = 1.5$ typical of a critical branching process. The average avalanche rate exhibits an inverse Omori law as a function of the time-to-failure, whereas the distribution of inter-avalanche times turns out to be consistent with the ETAS model of earthquake statistics. 
\end{abstract}

\maketitle

A wide range of disordered materials exhibit a common rheological response when loaded under creep conditions at constant stress levels below their short-time strength \cite{Deschanel2009_JPD,Heap2011_EPSL,Leocmach2014_PRL,Koivisto2016_PRE}: after a decelerating and a constant strain rate regime, deformation enters an accelerating regime where macroscopic failure is approached as a finite time singularity of the creep rate. Deformation proceeds in avalanches which reveal the discrete nature of plastic flow at the microscopic scale and the internal collective dynamics in the run-up to failure. In the present study we aim at linking the avalanche phenomena that are commonly observed in plasticity of disordered materials with the critical behavior associated with creep failure. We start out from  stochastic plasticity models that have been studied extensively in the recent literature \cite{Baret2002_PRL,Picard2004_EPJE,Talamali2012_CRM,Budrikis2013_PRE,Lin2014_PNAS,Lin2015_PRL,Liu2016_PRL,Sandfeld2015_JSTAT,Lin2016_PRX,Budrikis2017_NC} and generalize them to include thermal activation of deformation events in conjunction with local softening of the material as a consequence of plastic deformation \cite{Tuezes2017_IJF,Girard2010_JSTAT,Dahmen2009_PRL}. This provides us with a framework that allows to capture the characteristic feedback between local softening and increased rate of deformation which ultimately results in the nucleation and growth of a macroscopic shear band as the creep rate accelerates towards a finite time singularity. We use this framework to study universal signatures of the avalanche dynamics near failure.

The model we use coarse grains the microscopic details of plastic deformation events to represent the material as a 2D lattice of yielding elements \cite{Talamali2012_CRM,Budrikis2013_PRE,Sandfeld2015_JSTAT,Budrikis2017_NC}.  The state of an element $i$ centered at position $\Br_i$ is represented by (i) a local stress tensor $\boldsymbol{\Sigma}(\Br_i)$ which is the superposition of stresses resulting from external boundary conditions (for creep: temporally constant applied tractions) and internal stresses resulting from the heterogeneity of the plastic strain field, (ii) an accumulated plastic strain $\boldsymbol{\epsilon}(\Br_i)$ and (iii) a local yield threshold $\hat{\Sigma}(\Br_i)$ which characterizes the internal state of the element. Plastic deformation is governed by the yield function $\Phi =  \sqrt{(3/2)\boldsymbol{\Sigma{'}}:\boldsymbol{\Sigma{'}}} - \hat{\Sigma} = \Sigma_{\rm eq}  - \hat{\Sigma}$ where $\boldsymbol{\Sigma}{'}$ is the deviatoric part of the stress tensor $\boldsymbol{\Sigma}$ (for generalizations see \citet{Budrikis2017_NC}). $\boldsymbol{\Sigma}$ is computed from the external boundary conditions and the plastic strain field $\boldsymbol{\epsilon}(\Br_i)$ using standard Finite Element methodology.

Below the scale of resolution of our model, microscopic plastic re-arrangements ('deformation events') take place which on the element scale produce a tensorial plastic strain increment $ \Delta \boldsymbol{\epsilon} = \boldsymbol{\hat{\epsilon}}\Delta\epsilon $. The tensor $\boldsymbol{\hat{\epsilon}}$, which gives the 'direction' of the local strain increment,  is in the spirit of an associated flow rule chosen to maximize energy dissipation by setting $\hat{\epsilon}_{ij} = \partial \Phi/\partial \Sigma_{ij}$. 
Deformation events are activated according to the local yield function values $\Phi(\Br_i)$  using the following rules: (i) an event is activated instantaneously if $\Phi(\Br_i) > 0$; (ii) the duration of a deformation event is assumed negligibly small; (iii) if $\Phi(\Br_i) < 0$, an event is activated with finite rate that depends on temperature $T$ according to $\nu(\Br_i) = \nu_0 \exp [-E(\boldsymbol{\Sigma})/(k_{\rm B}T)]$ where $\nu_0$ is an attempt frequency within the element volume. We approximate the stress dependence of the activation energy $E$ by a linear dependency on the equivalent stress, $E = E_0 - V_{\rm A} \Sigma_{\rm eq}$ where $V_{\rm A}$ is an activation volume. The activation barrier goes to zero if $\Phi = 0$, hence $E_0 = \hat{\Sigma} V_{\rm A}$ and we can write the activation rate alternatively as $\nu(\Br_i) = \nu_{0} \exp [\Phi(\Br_i)/\Sigma_T]$ where the parameter $\Sigma_T = k_{\rm B} T/ V_{\rm A}$ characterizes the influence of thermal fluctuations on event activation. In the limit where the activation thresholds $\hat{\Sigma}$ are spatially uniform and the strain increments $\Delta \boldsymbol{\epsilon}$ are infinitesimally small, our model reduces for $T \to 0$ (no thermal effects) to a standard J2 plasticity model. On the other hand, at low stresses, the model reduces to a viscoplastic creep model where the rate of plastic flow is given by $\dot{\boldsymbol{\epsilon}} =  \Delta \boldsymbol{\epsilon} \nu_0 \exp(\Phi/\Sigma_T)$.

Statistical heterogeneity of the material is represented by considering the local yield thresholds $\hat{\Sigma}(\Br_i)$ as random variables which we assume to be Weibull distributed with exponent $k$ and mean value $\Lambda$, in line with recent molecular dynamics simulations \cite{Patinet2016_PRL} on glasses.  Evolution of the local thresholds due to internal structural changes within the elements is 
envisaged as a superposition of two processes: (i) after each deformation event, the local yield threshold is renewed, i.e., it is assigned a new random value from the Weibull distribution, hence the evolution of local thresholds proceeds in a stochastic manner; (ii) the mean threshold, and hence the scale parameter $\Lambda$, decreases exponentially as a function of the local strain, $\hat{\Sigma}_0(\Br) = \hat{\Sigma}_0 \exp\left(-\epsilon(\Br) f\right)$ where $\epsilon = \sqrt{(2/3) \boldsymbol{\epsilon}:\boldsymbol{\epsilon}}$ is the local equivalent plastic strain. 

Simulations are carried out under pure shear conditions by imposing on the free surfaces of the system spatially uniform tractions giving rise to a homogeneous external shear stress $\Sigma$ which is kept fixed during a simulation. Thermally activated deformation events are selected by the Kinetic Monte Carlo Method according to the stress-and temperature dependent local activation rates given above. Upon activation we increase the local strain at the activated site, re-compute the stress field and check whether, as a consequence of stress re-distribution, the condition $\Phi(\Br_i) < 0$ is fulfilled on any site. These sites also become activated and deform, leading to further stress changes and possible activation of further elements. The ensuing avalanche proceeds adiabatically as a series of deformation steps in each of which one or more elements are activated and yield (parallel update) until the inequality $\Phi(\Br_i) < 0$ is not fulfilled anywhere \cite{Budrikis2013_PRE,Budrikis2017_NC}. After termination of the avalanche we evaluate the avalanche size $S$ as the  total number of events activated during the avalanche and then make another Kinetic Monte Carlo step to determine the initiation site and initiation time of the next avalanche. We terminate the simulation as soon as a single avalanche induces a macroscopic average strain of 1, which ensures that this `infinite' avalanche is a clear outlier with respect to the prior avalanche statistics (see \figref{fig:avalfail} below). The starting time/strain of this avalanche is then identified as the failure time/strain. Unless otherwise stated, simulation results are averaged over many realizations of the stochastic evolution of local thresholds. 

\begin{figure}
	\includegraphics[width=0.50\textwidth]{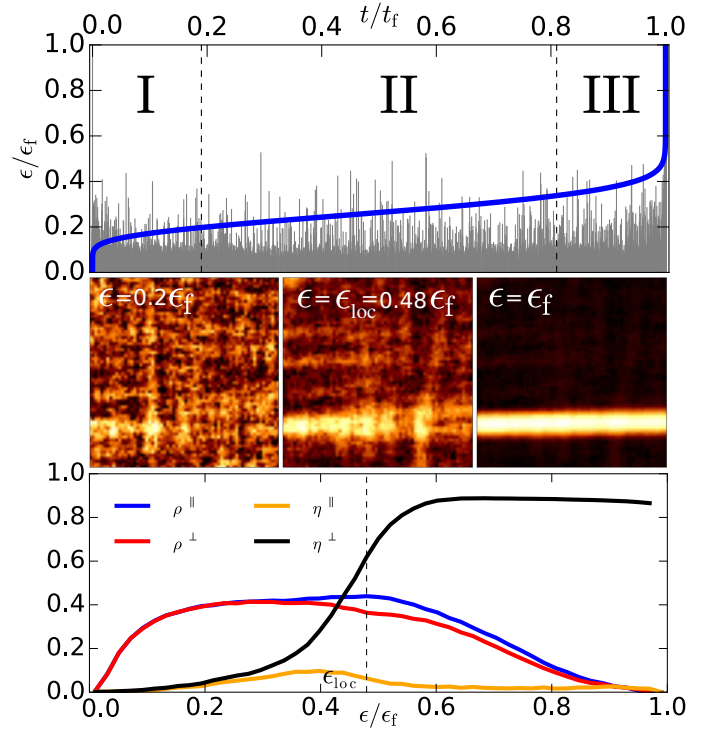}
	\caption{\label{fig:curve_patterns} Top: average creep-strain vs time curve for the default set of parameters, $\Sigma_T = 0.0075$;  background: strain increments in individual avalanches during a single realization; center: plastic strain localization patterns at different stages of the creep curve; bottom: strain evolution of the plastic event correlation and localization parameters; failure occurs, in system units, at $\epsilon_{f}=1.22$ and $t_{f}=1.73 \cdot 10^{12}$.}
\end{figure}

In the simulations stress is measured in units of $\hat{\Sigma}_{0}$, strain in units of $\hat{\Sigma}_{0}/E$ where $E$ is Young's modulus, and time in units of  $\nu_{0}^{-1}$. The model then depends on a non-dimensional coupling constant $C=E\Delta\epsilon/\hat{\Sigma}_{0}$ which controls the relative intensity of stress redistribution with respect to initial strength. The model relates to particular disordered systems through the values of $C$, $\hat{\Sigma}_{0}$, $k$, and $f$. Unless otherwise stated, we assume the default system parameters $C=0.05$, $L=64$, $k=4$, $f=0.1$. For the external stress we use a default value of $\Sigma = 0.7 \Sigma_{\rm c}$ where $\Sigma_{\rm c} = 0.323$ is the stress at which the default system fails immediately even in absence of thermal activation. The corresponding creep curves, shown in \figref{curve_patterns} (top) for $\Sigma_T = 0.0075$, exhibit typical three-stage behavior as observed in experiment \cite{Deschanel2009_JPD,Heap2011_EPSL,Leocmach2014_PRL,Koivisto2016_PRE}: An initial stage I of decelerating creep rates which is followed by an approximately linear stage II of constant creep rate and an accelerating stage III during which failure is approached as a finite-time singularity of the creep rate. Here we focus on this last stage of the creep curve and the approach to failure.  

Stage III is characterized by the emergence of a macroscopic shear band which breaks both the translational symmetry of the system and the symmetry between $x$ and $y$ directions (see \figref{curve_patterns}). To adequately capture the ensuing correlation structure, we introduce coordinates $\Br = (r_1,r_2)$ where the $r_1$ coordinate is parallel and the $r_2$ coordinate perpendicular to the mean direction of the final shear band. We then evaluate correlation coefficients between the locations $\Br$ and $\Br'$ of subsequent thermal activation events: $\rho^{\parallel}(\epsilon) = (\langle r_1 r^{\prime}_{1} \rangle_{\epsilon} - \langle r_{1} \rangle_{\epsilon}^{2})/\sigma(r_1,\epsilon)^2$, $\rho^{\perp}(\epsilon) = (\langle r_{2} r'_{2} \rangle_{\epsilon} - \langle r_{2} \rangle_{\epsilon}^{2})/\sigma(r_2,\epsilon)^2$. Here $\langle \cdot \rangle_{\epsilon}$ denotes an average over a narrow strain window centered at $\epsilon$ and $\sigma(r,\epsilon)$ is the standard deviation of event locations within that strain window. In addition we define localization coefficients $\eta_{\parallel} = 1-\sigma(r_1,\epsilon)^2/\sigma_0^2$ and $\eta_{\perp} = 1-\sigma(r_2,\epsilon)^2/\sigma_0^2$ where $\sigma_0$ is the standard deviation of locations that are distributed over the simulated sample in a completely random manner. Deformation events are initially statistically independent and homogeneously distributed, as reflected by near-zero correlation and localization coefficients. Correlations grow with time during the stationary creep regime (\figref{curve_patterns} (bottom)) as the elastic coupling favors correlated activation along directions where the internal stress created by an event is positive, leading to patterns typical of stochastic shear plasticity \cite{Talamali2012_CRM} (see pattern for $\epsilon = 0.2 \epsilon_{\rm f}$ in \figref{curve_patterns}). 

At the end of the linear creep regime, as deformation starts to accelerate towards failure, the slip event pattern exhibits a symmetry breaking transition where deformation localizes into a catastrophic slip band (patterns for $\epsilon = 0.5 \epsilon_{\rm f}$ and $\epsilon =\epsilon_{\rm f}$ in \figref{curve_patterns}). This is manifested by a sharp increase of the localization factor $\eta$. At the same time the correlation coefficients $\rho^{\perp}$ and $\rho^{\parallel}$ {\em decrease} showing that now events are localized in the slip band but the positions of subsequent events within the localization zone are not mutually correlated (\figref{curve_patterns} (bottom)). 

\begin{figure}
	\includegraphics[width=0.5\textwidth]{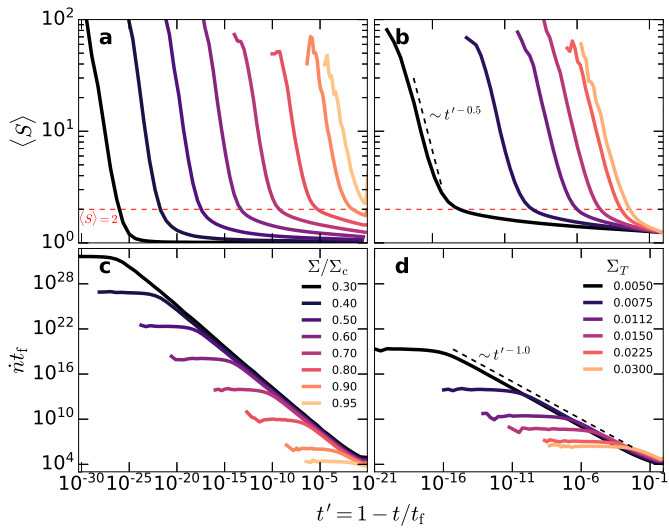}
	\caption{\label{fig:full_panel_rates} Top row (a,b): average avalanche size $\langle S \rangle$, bottom row (c,d): avalanche rate $\dot n$ rescaled by failure time $t_{f}$, both as functions of the reduced time-to-failure $t^{\prime}$; left column: data for varying external stress, right column: data for varying effective temperature.}
\end{figure}

\begin{figure}
\begin{tabular}{cc}
 	\includegraphics[width=0.26\textwidth]{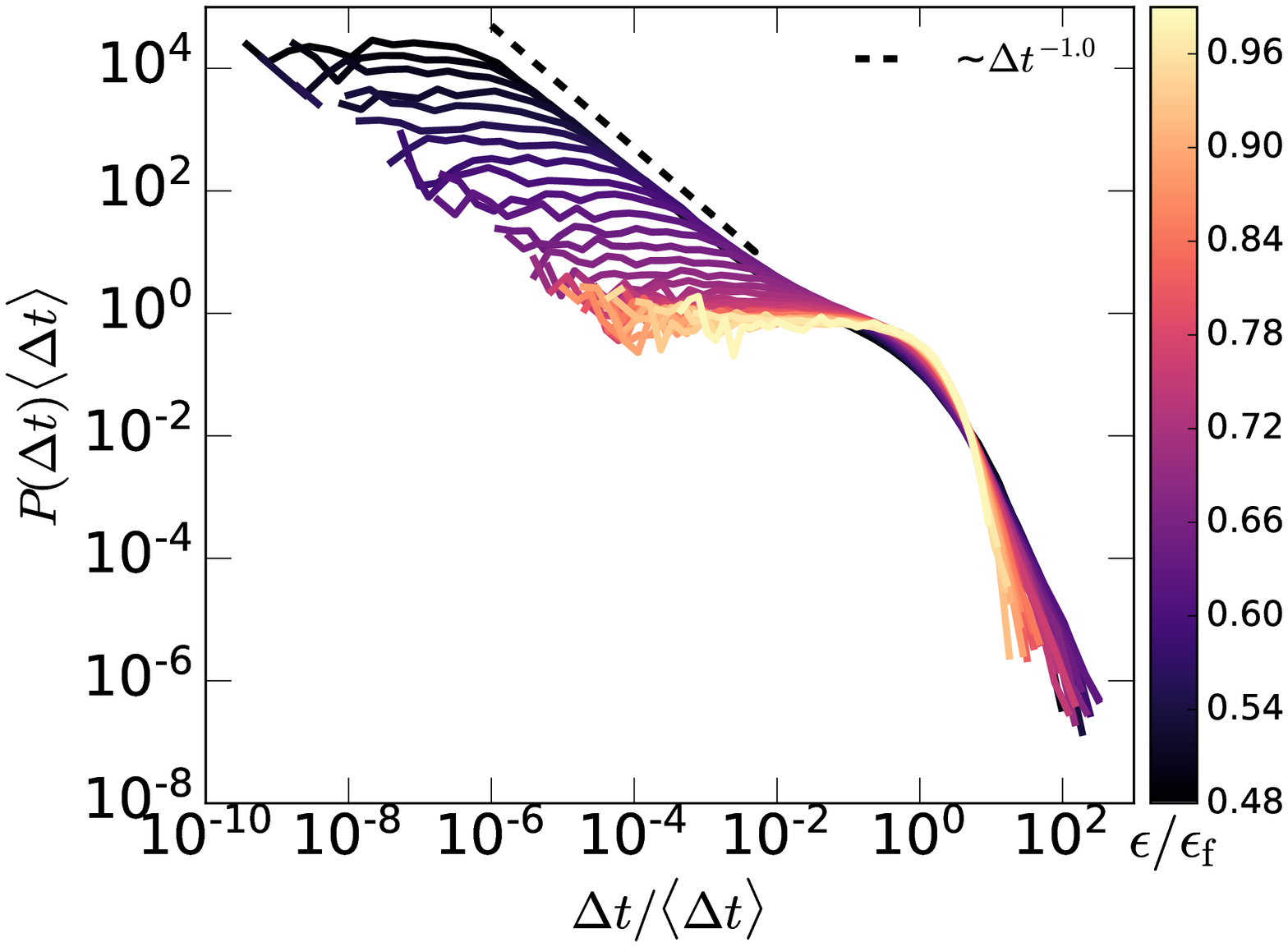}
	&
	\includegraphics[width=0.22\textwidth]{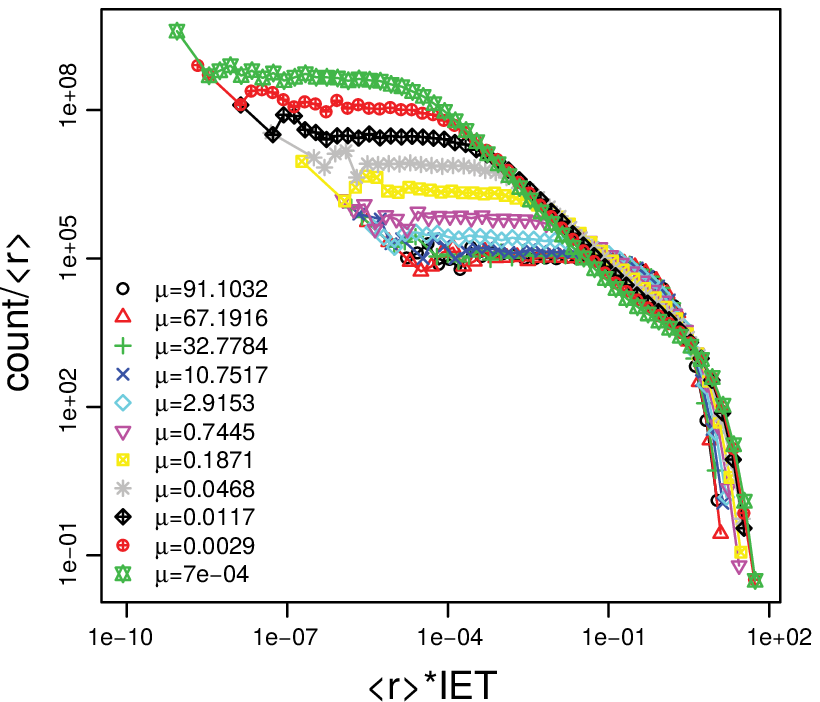}
	\end{tabular}
	\caption{\label{fig:avaltimestats} left: probability distribution $P(\Delta t)$ of inter-avalanche waiting times, in units of mean waiting time, for different strains (different avalanche rates); right: probability distribution of waiting times scaled by mean avalanche rate as obtained from ETAS model for different rates $\mu$ of spontaneously triggered avalanches, figure taken from \citet{Touati2009_PRL}.}
\end{figure}

We focus now on statistical signatures of the avalanche dynamics in Stage III. The strain rate is a function of avalanche size $\langle S \rangle$ and avalanche rate $\dot{n}$, both of which increase in the approach to failure, see \figref{full_panel_rates} which shows $\langle S \rangle$ and $\dot{n}$ as functions of the reduced time-to-failure, $t^{\prime} = 1-t/t_{f}$. Coincident with the localization of deformation, the system enters an {\em Omori regime} where the avalanche rate grows as a power of the reduced time-to-failure, $\dot n \propto {t^{\prime}}^{-p }$ where $p$ is close to 1. This precursor activity following inverse Omori's law is consistent with observations in geophysics and rock failure \cite{Leocmach2014_PRL,Lennartz-Sassinek2014_PRE, Schmid2012_JGR}. The correspondence between spatial localization of deformation activity and the beginning of Omori-like behavior is robust upon parameter variation. Over the Omori regime, the distribution of inter-event times undergoes qualitative changes as the event rate accelerates: Whereas at low event rates (at the beginning of creep stage III) the distribution has power-law characteristics, with increasing event rate close to the failure strain an approximately exponential distribution is approached (\figref{avaltimestats}, left). This behavior matches predictions derived by \citet{Touati2009_PRL} from the ETAS model of earthquake inter-event times (\figref{avaltimestats}, right). This finding is remarkable: ETAS is a stochastic model which aims at reproducing the phenomenology of earthquake time sequences in terms of a stochastic point process. Our model has a completely different structure -- it aims at a physical description of the spatio-temporal processes which control deformation localization and creep acceleration in materials failure. The fact that both models yield near -identical time sequences may serve as an indication that the present model of strain localization and activation captures essentials of the dynamics of earthquake faults.

\begin{figure}
	\includegraphics[width=0.5\textwidth]{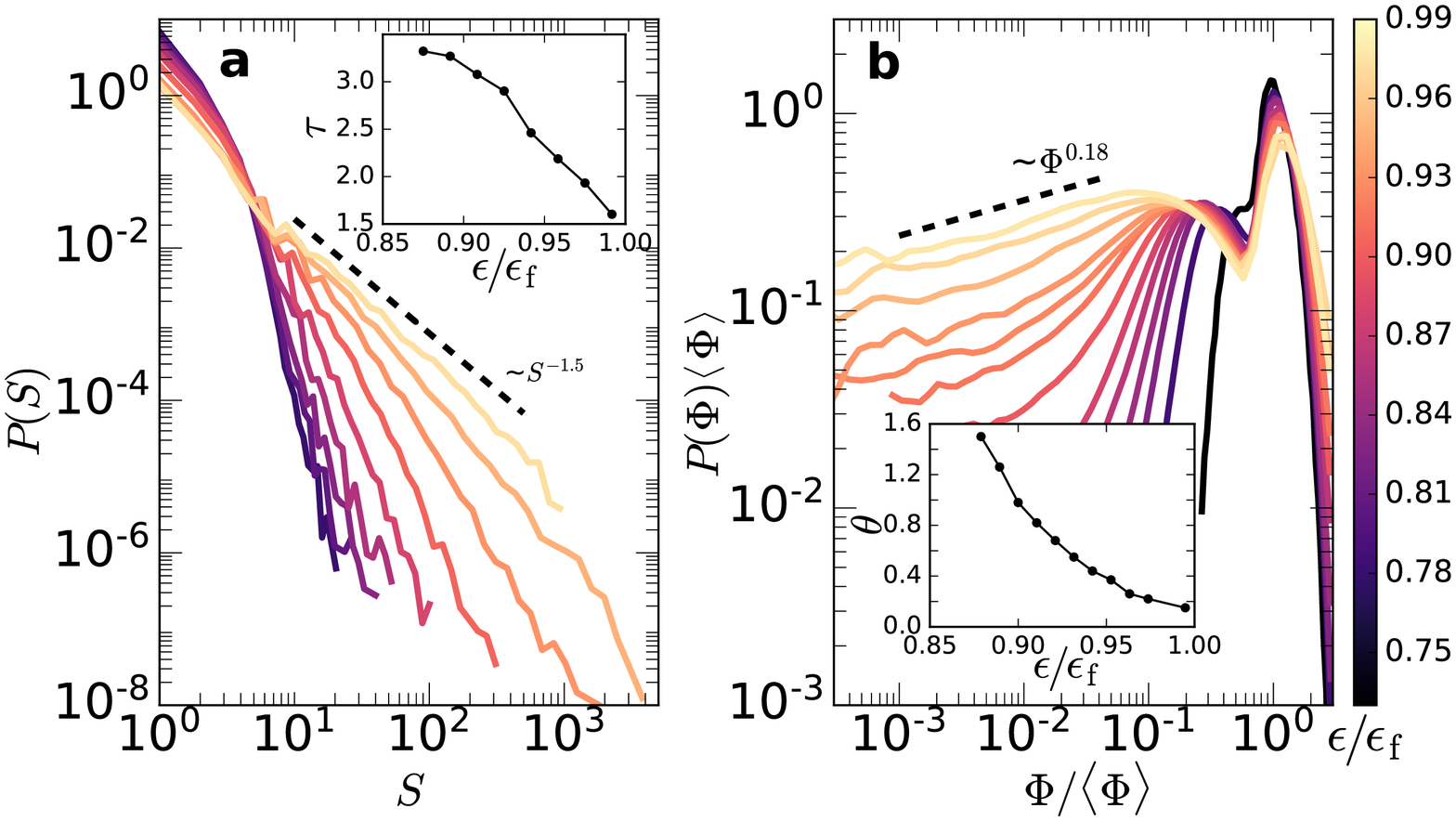}
	\caption{\label{fig:dist_soft} (a) Avalanche size distribution $P(S) \thicksim S ^{-\tau}$ at different strains, inset: $\tau$ vs. strain; (b) local stability distribution $P(\Phi) \thicksim \Phi^{\theta}$ at different strains, inset: $\theta$ vs. strain.}
\end{figure}

\begin{figure}
	\includegraphics[width=0.45\textwidth]{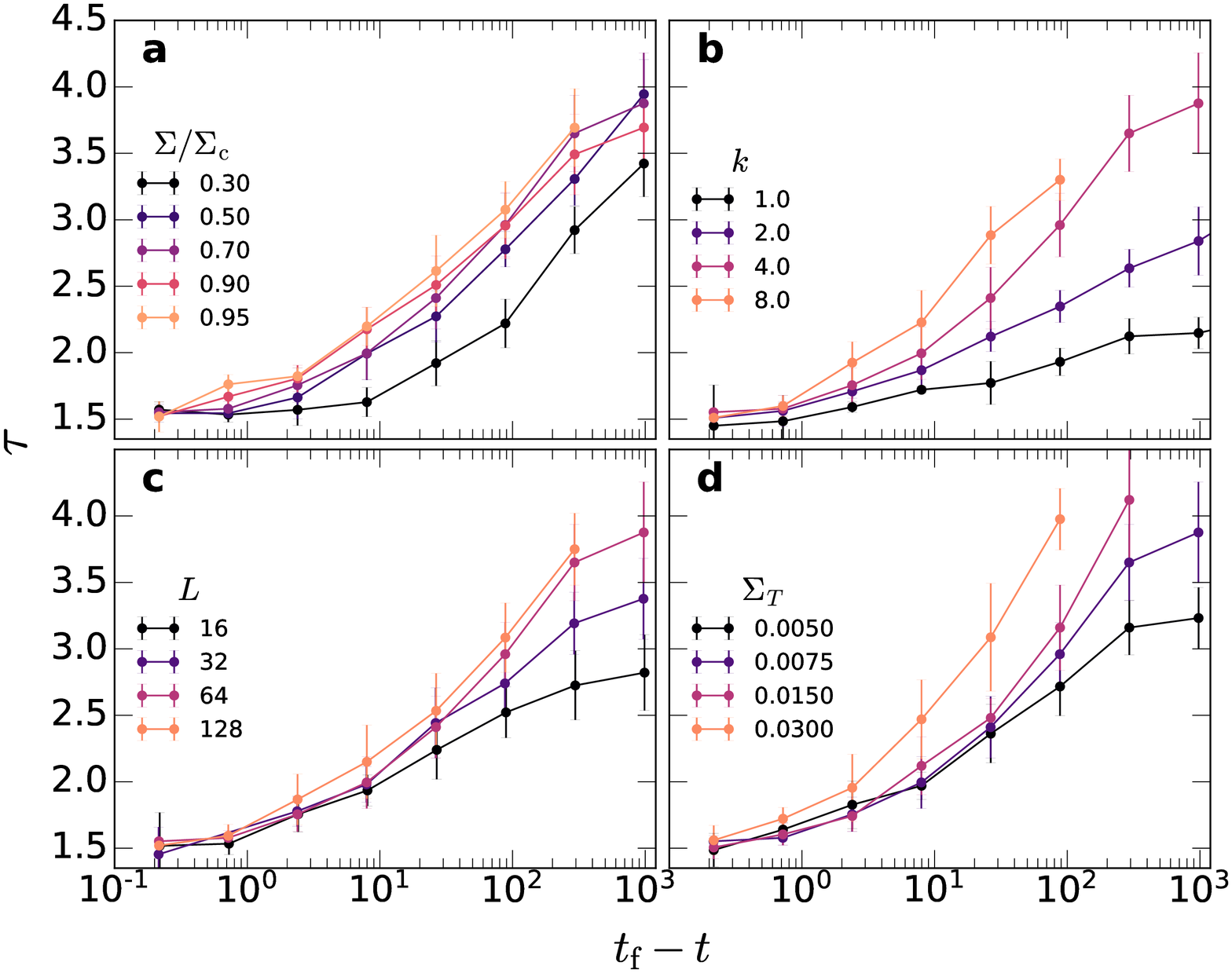}
	\caption{\label{fig:avalexponent}  Avalanche size exponent $\tau$ vs. time-to-failure; evolution depends on simulation parameters (external stress $\Sigma$, disorder $k$, system size $L$ and effective temperature $\Sigma_T$), exponents converge towards $\tau = 1.5$ at failure.}
\end{figure}

As the system approaches failure, the avalanche size distribution develops an extended power-law regime $P(S) \thicksim S ^{-\tau}$ (\figref{dist_soft} (a)). The exponent $\tau$ as determined by maximum likelihood estimation over different time windows is shown in \figref{avalexponent}; it decreases towards failure as observed in many experiments and geophysical contexts \cite{Lei2006,Soto-Parra2015_PRE,Jiang2016_AM,Amitrano2012_EPJS}. As $t \to t_{f}$, the values of $\tau$ converge from above to $\tau \rightarrow 3/2$ (see \figref{avalexponent}), a value that is typical of mean-field models envisaging avalanche dynamics as a critical branching process and matches the behavior found in in diverse stochastic models of failure, such as discrete element based \cite{Kun2013_PRE} or fiber-bundle models \cite{Pradhan2005_PRL}. Immediately before failure, the avalanche size distribution becomes independent of system and deformation parameters; the maximum avalanche size reached is controlled by system size and scales like $L^{d_f}$ with $d_f \approx 1.25$, \figref{avalfail}. This size scaling is in line with results reported earlier in the context of plasticity models, see e.g. \cite{Zaiser2007_JSTAT}.

\begin{figure}
	\includegraphics[width=0.4\textwidth]{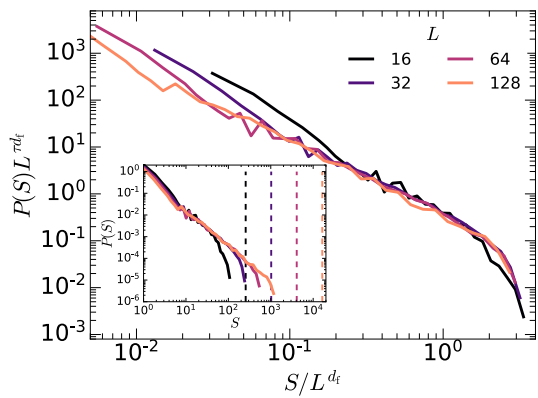}	
	\caption{\label{fig:avalfail} Avalanche size distributions near failure ($\epsilon/\epsilon_{f} = [0.995-1.0]$) for different system sizes $L$ rescaled with $L^{d_f}$ and $d_{f} = 1.25$; inset: original distributions, the dashed lines mark the sizes of the final avalanche at which the simulations are terminated.}
\end{figure}

Recent discussions of avalanche dynamics in plasticity in disordered media have focused on the statistics of a local stability index $x$, which is in scalar plasticity models defined as the difference between the local stress and the critical stress needed to activate deformation in a given elementary volume, see e.g. \cite{Lin2015_PRL,Lin2014_PNAS}. In the language of engineering mechanics, this stability index corresponds to the local value of the yield function $\Phi$ which provides a scalar measure of the distance of the (tensorial) stress state in a material volume element from the yield surface. In the present model, in the approach to failure the distribution $P(\Phi)$ becomes bimodal, \figref{dist_soft}(b), with a high-strength peak characterizing the plastically inactive region outside the shear band and a second peak at lower strength that corresponds to locations inside the shear banmd. The stability distribution inside the shear band exhibits power-law behavior for small $\Phi$, i.e., for local volumes that are close to the yield surface, $P(\Phi) \propto \Phi^{\theta}$. We observe that in the approach to failure, the exponent $\theta$ decreases and reaches small asymptotic values, see \figref{dist_soft}(b), inset. The stability exponent $\theta$ has been related to the avalanche exponent $\tau$ by \citet{Lin2015_PRL,Lin2014_PNAS}. The values $\theta \to 0$ and $\tau \to 1.5$, which we find asymptotically near failure, are characteristic of the classical mean field theory of avalanches in depinning transitions. At the same time we note that a simultaneous decrease of $\theta$ and $\tau$, as observed here, is inconsistent with the scaling relations of \citet{Lin2015_PRL,Lin2014_PNAS} which predict that a decrease of $\theta$ implies an  {\em increase} of $\tau$ and vice versa. 

Our model delineates a scenario of creep failure of a disordered material which matches the observations in many material and geosystems. In particular, the model captures the system-scale localization of deformation activity in a shear band which forms during the accelerating creep stage III and runs along the ultimate plane of failure, in line with experimental findings \citep{Lennartz-Sassinek2014_PRE,Jiang2016_AM,LeBouil2014_PRL,Renard2017_EPSL}. Interestingly, this system-scale localization goes along with a decreasing spatial correlation between sequential events {\em within} the slip  band. This prediction may be corroborated by detailed, spatially resolved AE analysis following the lines of \citet{Lennartz-Sassinek2014_PRE}. During the localization/acceleration stage, the global event rate exhibits an Omori-type acceleration towards failure. At the same time, the distribution of inter-event times shows with increasing global event rate a cross-over from power-law to exponential behavior which very accurately matches the predictions of ETAS type models \cite{Touati2009_PRL}, indicating a close link between intermittent behavior in softening-induced creep failure and earthquake statistics. The parallelism between the physically based model proposed here and a phenomenological stochastic rate model (ETAS) may be extremely useful for relating the phenomenological ETAS parameters to physical parameters controlling the deformation dynamics of different physical systems. Our findings have also interesting implications for avalanche statistics. As localization proceeds, we find that the avalanche exponent decreases near failure to a universal value close to $\tau = 1.5$. This time evolution of $\tau$ might explain the variability of experimentally determined $\tau$ values \cite{Lei2006,Soto-Parra2015_PRE,Jiang2016_AM,Amitrano2012_EPJS}. Finally, we note that generic relations
between the avalanche exponent and the local stability exponent $\theta$  \citep{Lin2015_PRL,Lin2014_PNAS}, which are supposed to hold in homogeneous systems, may be invalid in systems which are subject to stochastic activation in conjunction with damage and strain localization.

\bibliography{paper}

\end{document}